\documentclass[twocolumn]{aastex631}

\usepackage{amsmath}

\shorttitle{The QPO in RE\,J1034+396}
\shortauthors{Taylor et al.}

\begin{document}

\title{The QPO in RE\,J1034+396 originates in the hot corona}

\correspondingauthor{Chloe S. Taylor}
\email{ctaylor9@stanford.edu}

\author[0000-0003-3117-3476]{Chloe S. Taylor}
\affiliation{Department of Physics, Stanford University, 382 Via Pueblo Mall, Stanford, CA 94305, USA}
\affiliation{Kavli Institute for Particle Astrophysics \& Cosmology, Stanford University, 452 Lomita Mall, Stanford, CA 94305, USA}

\author[0000-0002-4794-5998]{Dan R. Wilkins}
\affiliation{Kavli Institute for Particle Astrophysics \& Cosmology, Stanford University, 452 Lomita Mall, Stanford, CA 94305, USA}
\email{dan.wilkins@stanford.edu}

\author[0000-0003-0667-5941]{Steven W. Allen}
\affiliation{Department of Physics, Stanford University, 382 Via Pueblo Mall, Stanford, CA 94305, USA}
\affiliation{Kavli Institute for Particle Astrophysics \& Cosmology, Stanford University, 452 Lomita Mall, Stanford, CA 94305, USA}
\affiliation{SLAC National Accelerator Laboratory,  2575 Sand Hill Road, Menlo Park, CA 94025, USA}
\email{swa@stanford.edu}

%% Mark off the abstract in the ``abstract'' environment. 
\begin{abstract}

RE\,J1034+396 is one of the few active galactic nuclei with a significant quasi-periodic oscillation (QPO).  The QPO has been observed in over 1 Ms of \textit{XMM-Newton} observations spanning over a decade.  We investigate the power spectral density function (PSD) of 7 long ($\sim$90 ks) \textit{XMM-Newton} observations of the active galactic nucleus RE\,J1034+396 in two energy bands.  The soft (0.3-0.5 keV) band targets emission from the disk, while the hard (2-7 keV) band isolates the primary X-ray continuum emission from the corona.  The QPO is significantly detected in the hard band of 5 of the 7 observations.  The best fitting models indicate that the QPO detection in both bands is entirely attributable to the coronal emission with no additional contribution from the disk.  This explains the strong coherence between the hard and soft bands at the QPO frequency.  The covariance spectrum is consistent with this picture as the variability at QPO frequencies is attributed solely to fluctuations in the hot corona.  The time lag as a function of energy is well described by a $\sim$2000 s intrinsic soft lag, resulting from the disk responding to emission from the corona, that undergoes phase wrapping at approximately the QPO frequency.  By demonstrating that in this system the QPO arises in the corona, we provide new insights into the mechanisms generating QPOs.

\end{abstract}

\section{Introduction} \label{sec:intro}
\noindent Black hole X-ray binaries (BHBs) and active galactic nuclei (AGN) are powered by accretion onto black holes.  This accretion process is expected to be dependent on the mass and spin of the black hole \citep{accr1,accr2}.  Quasi-periodic oscillations (QPOs) are commonly observed in the X-ray flux of BHBs, but much is unknown about the physical mechanisms that generate them.  BHBs ($M_\textrm{BH}\sim 10 M_\odot$) exhibit both low frequency ($\lesssim30$ Hz) and high frequency ($\gtrsim60$ Hz) QPOs in their power spectra \citep{xrbrev}.  Of these, high frequency (HF) QPOs are relatively rare and occur in systems with high mass accretion rates \citep{moreqpo1,moreqpo2}.  The periods of HF QPOs correspond to timescales associated with the inner accretion disk.  Due to the expected scale invariance of accretion, we expect to find HF QPOs in AGN as well.

The first convincing detection of an AGN QPO was the $\sim2.6 \times 10^{-4}$ Hz QPO in the Seyfert galaxy RE\,J1034+396 \citep{rejqpo1}.  Given the high mass accretion rate of RE\,J1034+396 (L/L$_{Edd}$ $\sim$ 1) and a linear mass scaling \citep{massscale} the QPO is believed to be an analog of the HF QPOs found in BHBs \citep{QPOana}.  Measurements of the stellar bulge velocity dispersion and second moment of the H$\beta$ line estimate the mass of the black hole in RE\,J1034+396 to be $\sim1-4 \times 10^{6}$ \(M_\odot\) \citep{goodM}.  Assuming a linear mass scaling, the QPO in RE\,J1034+396 should correspond to a $\sim26-104$ Hz QPO in a $10 M_\odot$ BHB.

RE\,J1034+396 has been the subject of deep, follow-up observations since the discovery of the QPO in a 90 ks \textit{XMM Newton} observation in 2007.  \citet{fourQPO} detect the QPO in only the hard (1-4 keV) band in four observations totaling $\sim$160 ks taken in 2010 and 2011.  This led to the suggestion that the QPO may be generated in the hot corona that dominates above $\sim$ 1 keV.  An additional 70 ks observation in 2018 detected the QPO in both soft (0.3-1 keV) and hard (1-4 keV) bands \citep{2018qpo}.  Interestingly, the hard band lagged the soft band by $\sim 400$ s with high coherence, leading \citet{2018qpo} to argue that the QPO originates in the soft component.  A final $\sim$ 1 Ms of observations taken between November 2020 and May 2021 revealed a QPO in the hard band of all 10 observations, while 4 observations showed no significant detection in the soft band \citep{newestqpo}.  These observations revealed two time lag reversals and suggest that the lag tracks QPO frequency.

Spectral analysis of RE\,J1034+396 reveals a large soft X-ray excess.  The soft excess is believed to arise primarily from repeated Compton scattering of thermal photons from the accretion disk by a warm plasma spanning the surface of the inner disk \citep{soft,soft2}.  This suggests that the warm plasma is an extension of the disk, allowing us to use the soft X-ray emission to probe the disk's properties.  Future work by Yu et al. (in prep) shows that the timing properties of the soft excess directly associate with the reflection from the inner disk, reinforcing this picture.  The soft disk emission dominates at low energies ($< 0.5$ keV).  At high energies ($> 2$ keV), the hot coronal power law like emission dominates.  Figure~\ref{fig:dia} presents a schematic illustrating the origin of the hard and soft emission.  We can take advantage of this clear separation to determine where the QPO originates.

%We revisit the most recent publicly available \textit{XMM Newton} observations and investigate the isolated disk (0.3-0.5 keV) and corona (2-7 keV) emission.  We evaluate the significance of the QPO in these observations and use Fourier analysis techniques to determine the origin of the QPO.

\section{Observations and Data Reduction}

\begin{figure}[htbp]
\centering
\includegraphics[width=\columnwidth]{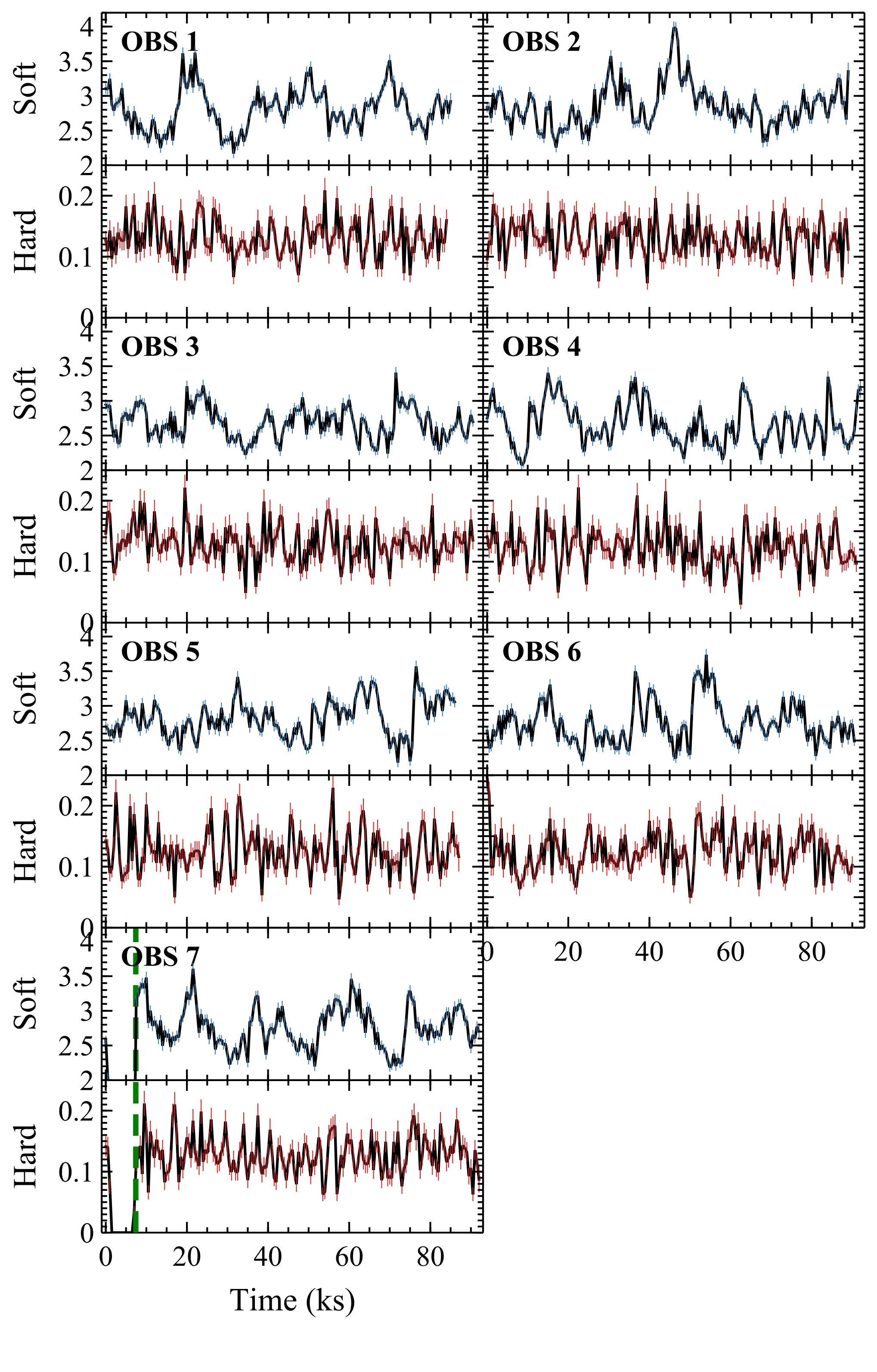}
\caption{The light curves for the 7 observations in Table~\ref{tab:obs} in the soft (0.3-0.5 keV) and hard (2-7 keV) bands.  The data is binned to 500 s with error bars, with error bars in blue (soft) and red (hard).  The light curve in Obs 7 shows zero counts when affected by background flaring.  This section of the light curve (before the vertical green dashed line) was trimmed for all analyses in this work.}
\label{fig:lc}
\end{figure}

\noindent We use seven of the 10 publicly available observations made with \textit{XMM-Newton} \citep{XMM} of RE\,J1034+396 taken between 2021 April 26 and 2021 May 31 (see Table~\ref{tab:obs} and Fig.~\ref{fig:lc}).  We exclude 3 of the 10 observations due to substantial contamination by an additional highly variable background due to soft proton flaring.  We analyze data from the EPIC-pn camera taken in `small window' mode \citep{PN}.  The observations were reduced using the XMM \textsc{science analysis system} (SAS) v21.0.0.  Event lists from the EPIC-pn camera were reprocessed and filtered using the \textsc{epproc} task and the latest calibration files.  Source light curves were extracted from a 35 arcsec region centered on the AGN and the background was taken from a different 35 arcsec region.  Light curves were constructed from the event lists using \textsc{evselect} with filter conditions \textsc{pattern} = 0-4 and \textsc{flag} = 0.  We used the \textsc{epiclccorr} task to correct light curves by accounting for dead time and exposure variations.  The light curves were binned to 100 s.

\begin{table}[htbp]
    \centering
    \caption{Summary of the \textit{XMM-Newton} observations used in this analysis.  The columns list: (1) observation number; (2) observation ID; (3) start date of the observation; and (4) duration of the flare-subtracted contiguous segment of the EPIC-pn observation (rounded down to the nearest ks). }
    \label{tab:obs}
    \begin{tabular}{cccc}
        \hline
        Obs No. & Obs ID & Date (UT) & Duration (ks) \\
        (1) & (2) & (3) & (4) \\
        \hline
        \hline
        1 & 0865011001 & 2020-11-30 & 85\\
        2 & 0865011101 & 2020-12-04 & 89\\
        3 & 0865011201 & 2020-12-02 & 91\\
        4 & 0865011301 & 2021-04-24 & 92\\
        5 & 0865011401 & 2021-05-02 & 87\\
        6 & 0865011501 & 2021-05-08 & 91\\
        7 & 0865011801 & 2021-05-30 & 85\\
        \hline
    \end{tabular}
\end{table}

\section{Power Spectrum}
\begin{figure}[htbp]
\centering
\includegraphics[width=\columnwidth]{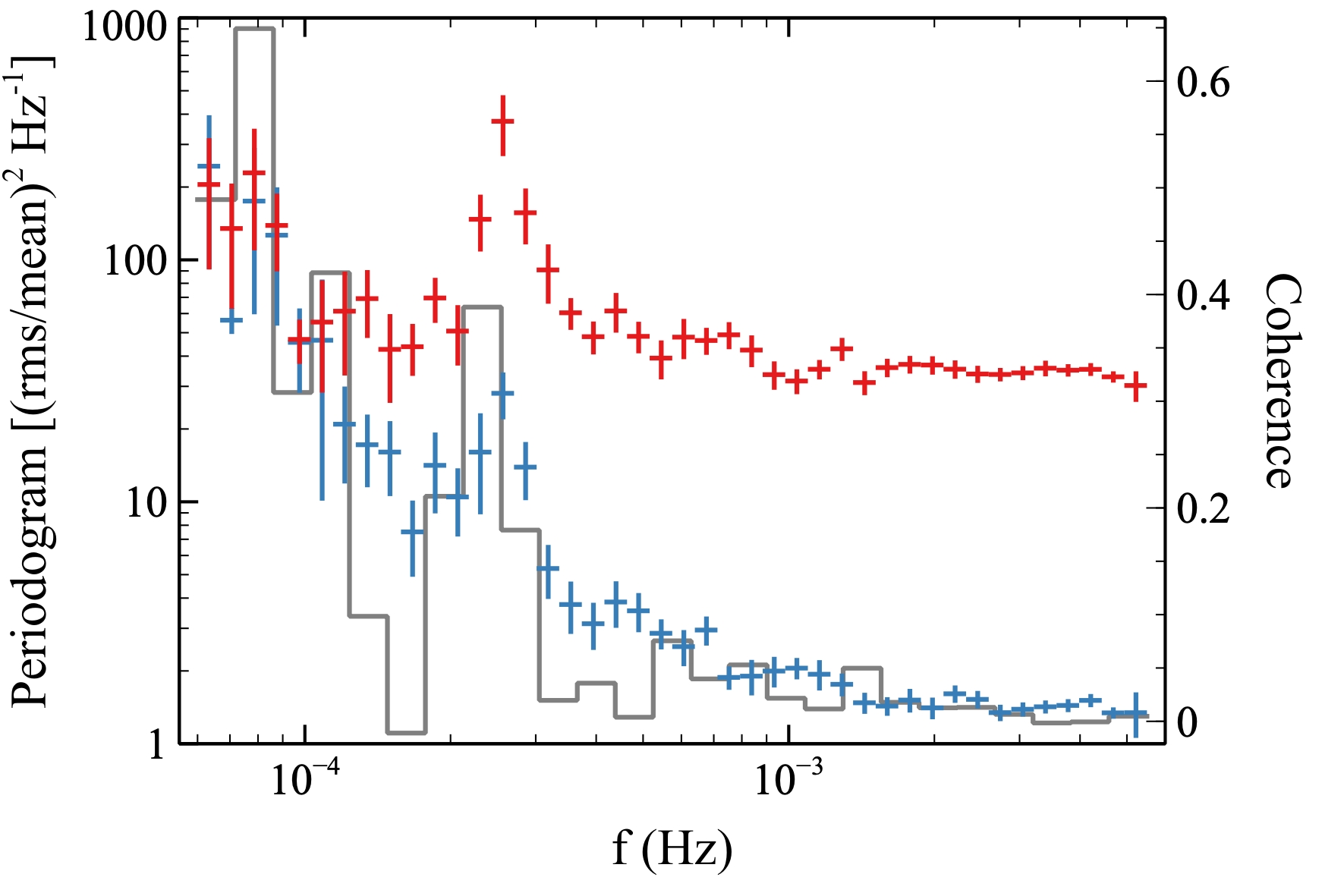}
\caption{The stacked peiodogram of all 7 observations in the hard (red) and soft (blue) bands.  The coherence is shown in grey.}
\label{fig:pscoh}
\end{figure}

\noindent The power spectral density function (PSD) was estimated from the periodogram using an [rms/mean]$^2$ normalization \citep{psd}.  We estimated the periodogram in two bands; 0.3-0.5 keV (soft) and 2-7 keV (hard).  The soft band was chosen so that the Comptonized disk component dominates and contains roughly $\sim 90\%$ disk emission and $\sim 10\%$ corona emission, while the hard band was chosen to isolate the power law like component from the hot corona (see Fig.~\ref{fig:dia} for a schematic detailing the origin of the hard and soft emission).  These fractions are established from the detailed spectral analysis of RE\,J1034+396 reported by \citet{rejUFO}.  The stacked periodograms of all seven observations in both the hard (red) and soft (blue) bands are shown in Fig.\ref{fig:pscoh}.  While the stacked periodogram is shown for the sake of clarity, all power spectral analyses were performed on the individual power spectra of each observation.

\begin{figure}[htbp]
\centering
\includegraphics[width=\columnwidth]{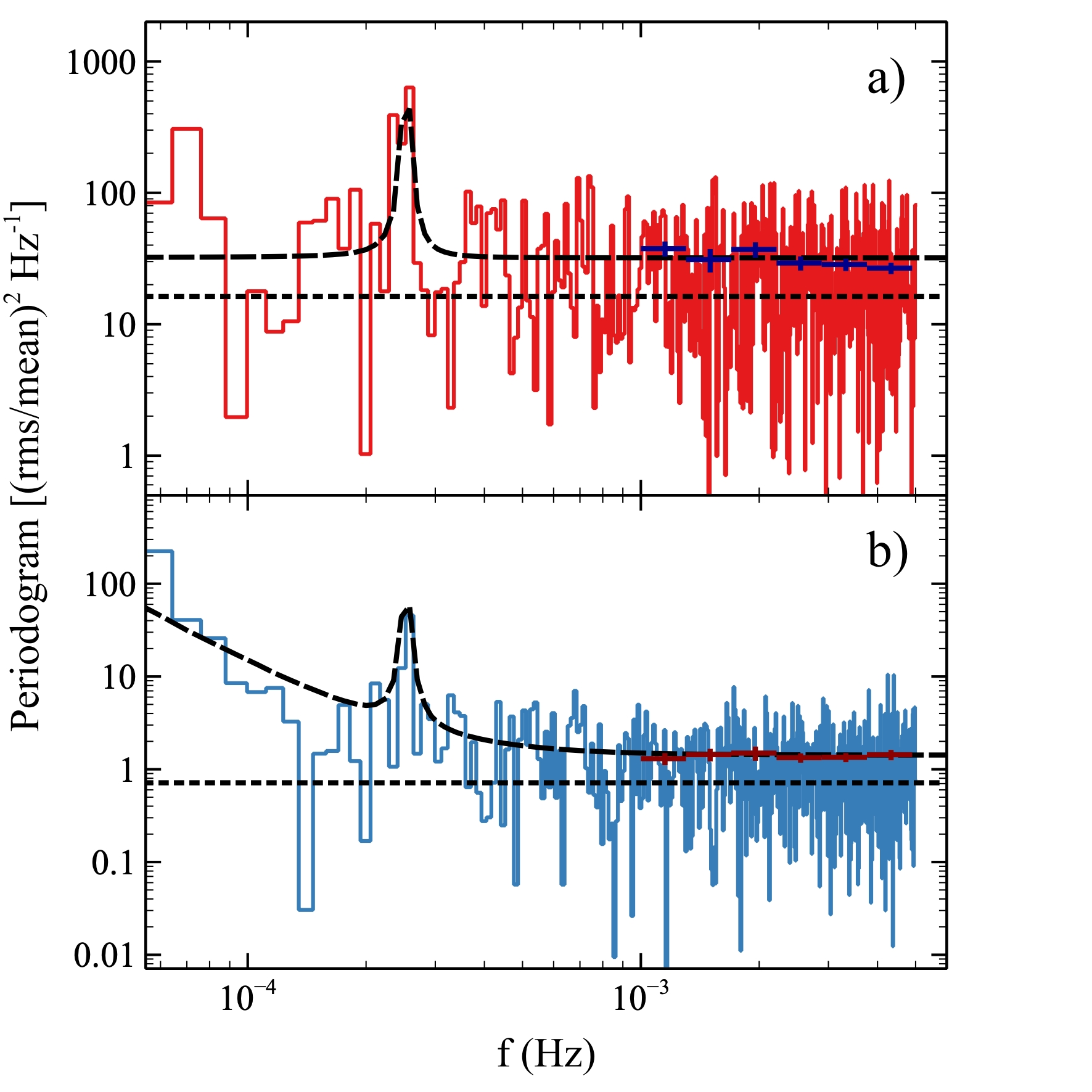}
\caption{(a) The hard band periodogram of Obs 1.  The best fitting model (Model A) is shown in dashed black.  The Poisson noise is indicated in dotted black.  (b) The soft band periodogram of Obs 1.  The best fitting model (Model A) is shown in dashed black.  For visualization purposes, above $f>10^{-3}$ Hz we overlay the binned data in six log spaced bins.  The Poisson noise is indicated in dotted black and reflects the expected level from counting statistics rather than the model fit.  We note that the model sits above the data because the Whittle likelihood assumes that the data are drawn from an exponential distribution.}
\label{fig:pscoh2}
\end{figure}

The unbinned power spectra for each observation were fit using the maximum likelihood method described in \citet{maxlik}.  We first explored two simple continuum models; a constant (Model 1)
\begin{equation}
    P(\nu) = C
\end{equation} modeling Poisson noise; and a power law plus constant (Model 2)
\begin{equation}
    P(\nu) = N\nu^{-\alpha} + C
\end{equation}
with normalization $N$.  We use the Whittle likelihood to approximate the likelihood function \citep{maxlik} and the Bayesian Information Criterion (BIC) for model selection, adopting a threshold of $\Delta\textrm{BIC}>10$ for selection of Model 2 over Model 1 \citep{BIC}.  Model 1 (constant) was preferred for all but one of the observations in the hard band (see Table~\ref{tab:info} Column 2).  Model 2 (constant plus power law) was preferred for all observations in the soft band, reducing the BIC by more than 100 in each case.  We note that we expect the stochastic variability of the corona to follow a red noise spectrum, so when Model 1 is preferred the continuum variability is dominated by white noise.  This white noise is the Poisson noise inherent in the measurement, which fits the hard spectrum well (see the black dotted lines in Fig.~\ref{fig:pscoh2}).

%The low frequency power law component is most significant in Obs 7.  In this observation, the change in Whittle likelihood between Model 1 and Model 2 in the soft band is $\sim 800$, as opposed to $\sim 100-300$ in the other observations.  We note that this is the same observation where \citet{rejUFO} observed the spectral signature of an ultrafast outflow (UFO) acceleration, hinting at a connection between low frequency variability and the mechanism responsible for lifting disk material before it is accelerated to relativistic velocities.

Using the preferred continuum model per observation, we simultaneously fit the hard and soft periodograms to explore two QPO models.  The first is a tied Lorentzian (Model A)
\begin{equation}
    P_{H}(\nu) = con. + L_1; \hspace{1mm} 
    P_{S}(\nu) = con. + \frac{F_\mathrm{PL}}{F_\mathrm{comp}}L_1,
    \tag{A}
\end{equation}
where $con.$ represents the preferred continuum model, $\frac{F_\mathrm{PL}}{F_\mathrm{comp}}$ is the ratio of coronal power law flux to the Comptonized disk flux in the soft band, and $L_1$ is the Lorentzian.  $\frac{F_\mathrm{PL}}{F_\mathrm{comp}}$ is set by the spectral fits performed in \citet{rejUFO}.  Model A assumes the QPO originates entirely in the corona and only appears in the soft band due to the modest contribution from the hot coronal continuum emission at low energies.  The second model (Model B)
\begin{equation}
    P_{H}(\nu) = con. + L_1; \hspace{1mm} 
    P_{S}(\nu) = con. + \frac{F_\mathrm{PL}}{F_\mathrm{comp}}L_1 + L_2
    \tag{B}
\end{equation}
introduces a second Lorentzian, $L_2$, allowing for additional, separate QPO power in the disk emission.  We again use the BIC to select between these models, and determine the improvement each provides over the continuum model alone.

The BIC favors Model A over the continuum model in 5 of the 7 observations (see Table~\ref{tab:info}), i.e. five of the seven observations provide QPO detections.  We tested the significance of the five QPO detections by simulating $5000$ soft band light curves and $5000$ hard band light curves using \textsc{pylag} following the method of \citet{LCsim}.  The simulated light curves had the same number of bins, mean count rate, and variance as the light curve in Obs 1.  We note that the maximum significance of this method is $3.5\sigma$ due to the limited number of simulated light curves.  None of the power spectra of the simulated light curves reduced the BIC by $>10$ for Model A, with Lorentzian centroid constrained within the QPO frequency range $(2.4-2.8)\times10^{-4}$, compared to the continuum model.  This means that the five BIC selected QPO detections (Obs. 1-5) are highly significant at $>3.5 \sigma$.  One of the two observations where the continuum model is preferred is the one with the low frequency power law in the hard band.  The low frequency power law hides some of the QPO signal in this observation, making the QPO detection less significant.  There are no observations where Model B is preferred to Model A.  Thus, there is no evidence for QPO power originating in the disk.  The best fitting model (Model A) applied to the hard and soft periodogram of Obs 1 is shown in Fig.~\ref{fig:pscoh2}.  We conclude that the QPO arises in the corona with no significant contribution from the disk.  Table~\ref{tab:info} lists the continuum and QPO properties for all seven observations.

Although Model B is not preferred for any observation, indicating no evidence for QPO power originating in the disk, we can still use it to place limits on possible disk contribution to the QPO.  The best fitting disk contribution to the QPO power is consistent with $0\%$ at the $1\sigma$ level across all 7 observations, with a $99.9\%$ upper limit of $73\%$.  We note that we are unable to provide tight constraints because the fit is affected by only a few data points.

\begin{table*}[htbp]
    \centering
    \caption{Results from Fourier analysis of RE\,J1034+396.  The columns list: the observation number; whether a power law is required to fit the hard band PSD continuum; the change in the Whittle likelihood between the continuum model and Model A; the best fitting Model A parameters (QPO frequency, QPO width, and hard band QPO normalization); the ratio of coronal power law flux to Comptonized disk flux in the soft band; whether we measure a hard or soft lag at the QPO frequency; and the measured time lag at the QPO frequency.}
    \label{tab:info}
    \begin{tabular}{ccccccccc}
        \hline
        Obs No. & Power law & $\Delta$Whittle & QPO freq. & QPO width & 
        QPO norm & F$_\textrm{PL}$/F$_\textrm{comp}$ & H/S Lag & QPO Lag \\
         &  &  & ($10^{-4}$ Hz) & ($10^{-5}$) & ($10^{-4}$) &  &  & (s) \\
        \hline
        \hline
        1 & No & 31 & $2.53^{+0.02}_{-0.03}$ & $0.44^{+0.5}_{-0.3}$ & $6.04^{+16.84}_{-4.26}$ & $0.137^{+0.006}_{-0.009}$ & S & $256\pm60$  \\
        2 & No & 28 & $2.57^{+0.07}_{-0.08}$ & $2.18^{+2.04}_{-1.09}$ & $1.93^{+3.07}_{-1.34}$ & $0.132^{+0.005}_{-0.010}$ & S & $598\pm140$ \\
        3 & No & 24 & $2.53^{+0.01}_{-0.03}$ & $0.28^{+0.53}_{-0.24}$ & $0.11^{+0.80}_{-0.10}$ & $0.122^{+0.010}_{-0.011}$ & S & $591\pm85$ \\
        4 & No & 37 & $2.81^{+0.07}_{-0.06}$ & $2.08^{+2.09}_{-1.01}$ & $4.42^{+7.52}_{-2.96}$ & $0.132^{+0.011}_{-0.005}$ & H & $882\pm89$ \\
        5 & No & 67 & $2.67^{+0.08}_{-0.08}$ & $3.11^{+2.42}_{-1.42}$ & $11.97^{+12.86}_{-6.85}$ & $0.135^{+0.008}_{-0.005}$ & H & $867\pm60$ \\
        6 & Yes & 13 & $2.64^{+0.06}_{-0.06}$ & $1.35^{+2.29}_{-0.85}$ & $0.86^{+2.31}_{-0.73}$ & $0.120^{+0.011}_{-0.009}$ & H & $914\pm98$ \\
        7 & No & 15 & $2.31^{+0.18}_{-0.06}$ & $3.16^{+12.69}_{-1.27}$ & $0.77^{+2.63}_{-0.59}$ & $0.130^{+0.011}_{-0.009}$ & S & $515\pm103$ \\
        \hline
    \end{tabular}
\end{table*}

\section{Variability at the QPO Frequency}
\noindent We use the cross spectrum to learn about the frequency dependent variability in RE\,J1034+396.  We compute the cross spectra for each observation using \textsc{pylag}, following Equation (9) in \citet{psd}.

\subsection{Coherence}
We use the binned cross spectrum to calculate the coherence $\gamma^2$.  The coherence measures the linear correlation between two light curves and is defined  in the range [0,1] \citep{coh}.  In this case, we are interested in the fraction of the total variability that can be predicted by a linear transformation of the hard band.  A binned cross spectrum is necessary to cancel uncorrelated noise in the coherence calculation; otherwise, the coherence will always be unity.

We calculate the cross spectrum from the Fourier transforms of each continuous observation segment.  We average across all of the observations in each frequency bin to obtain the stacked coherence between the hard and soft bands in 25 log spaced bins between $6\times10^{-5}$ Hz and $5.5\times10^{-3}$ Hz.  We note that the observed coherence remains consistent across observations, allowing us to use the stacked coherence.   As is typical in Seyferts, the coherence in RE\,J1034+396 is high at low frequencies and decreases with increasing frequency until Poisson noise dominates \citep{cohpat}.  We observe a significant peak in coherence at the QPO frequency (Fig.~\ref{fig:pscoh}a).  The strong coherence indicates that the QPO variability between energy bands is highly correlated.  This is unsurprising given our finding that the QPO variability in the soft band can be entirely attributed to contamination from the hot corona.

\subsection{Time Lags}
\begin{figure}[htbp]
\centering
\includegraphics[width=\columnwidth]{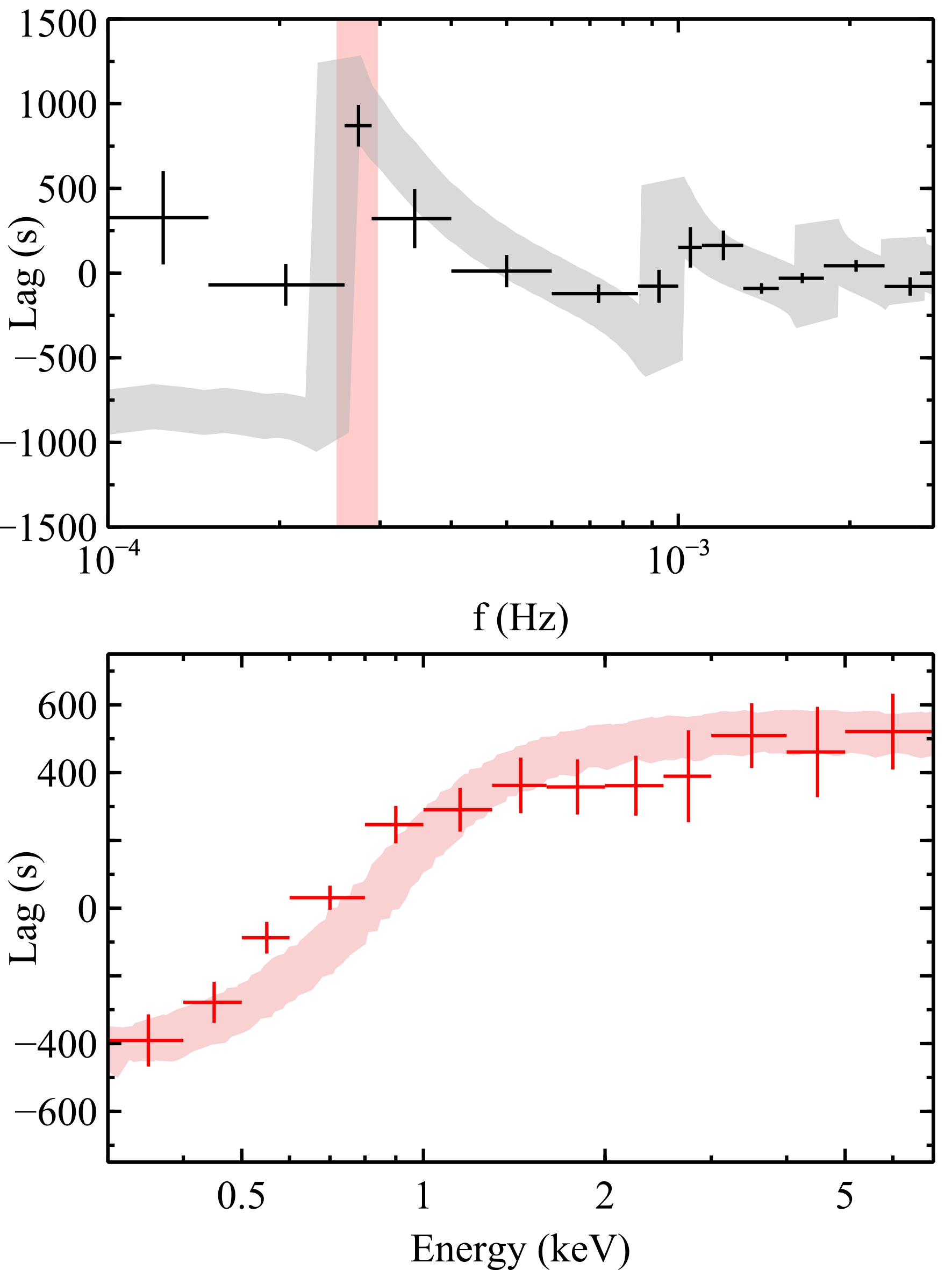}
\caption{a) The stacked lag frequency spectrum of all seven observations.  Predicted phase wrapping due to a diluted $\sim2000$ s soft lag is shown in light grey.  A typical low frequency hard lag dominates below $\sim2\times10^{-4}$ Hz.  Above this frequency, the phase wrapped soft lag is a good description of the data.  b) The stacked lag energy spectrum of all seven observations at the QPO frequency ($2.6-2.9\times10^{-4}$, red band in above figure).  We note that this frequency range excludes the measured QPO frequencies of Obs 1-3 and 7.  This avoids erasure of the shape of the stacked lag energy spectrum due to phase wrapping at $\sim2.6\times10^{-4}$ Hz.  The predicted lag energy spectrum where the soft Comptonized disk emission lags the continuum emission from the corona is shown in light red.  This appears as a hard lag due to phase wrapping.}
\label{fig:lf}
\end{figure}

We can also estimate the phase lag ($\Delta\phi$) between the hard and soft bands at each frequency in the cross spectrum.  This phase lag can be converted into a time lag ($\Delta\tau=\Delta\phi/2\pi f$).  We use the coherence to estimate error bars on the lag-frequency spectrum as outlined in \citet{psd} Equation (12).  Fig.~\ref{fig:lf}a shows the stacked lag frequency spectrum following the convention that a positive time lag indicates that the hard band lags the soft band.  We again note that the observed lag-energy spectra are consistent across all observations, justifying the use of the stacked spectrum.  A low frequency hard lag typically attributed to the propagation of accretion rate fluctuations is seen in RE\,J1034+396 \citep{lagfreq1,lagfreq2}.  

Measured phase lags can be affected by a variety of factors, including phase wrapping and dilution.  Phase wrapping occurs at the frequency $f$, where the intrinsic lag between the driving continuum and the reprocessed emission is $f=\frac{1}{2\Delta\tau}$.  At this frequency, the wave could have been shifted forward or backward by half a wave.  This results in a `phase wrap' that causes the measured lag to become negative.  Dilution is caused by the presence of the driving continuum in the reflecting band.  The main effect of dilution is to reduce the amplitude of the measured lag.  The reduction in measured lag depends on a constant $R$ that we define as the relative amplitude of the reflection relative to the diluting driving continuum.  Dilution reduces the observed lag by a factor of $R/(1+R)$.  Above $\sim2\times10^{-4}$ Hz, the lag frequency spectrum of RE\,J1034+396 is well described by a phase wrapped $\sim 2000$ s soft lag with a dilution $R\sim1$ (Fig.~\ref{fig:lf}a).  Although the Comptonized disk contributes $\sim90\%$  of the flux in the soft band, the amplitude of the disk variability is comparable to the amplitude of the coronal variability in this band, resulting in a dilution of $R\sim1$.  We find that the soft disk emission is responding to the hard corona continuum emission.  We see multiple measured lag sign changes at increasing frequencies as predicted by a phase wrap model.

The phase wrapping interpretation explains the time lag reversals discovered by \citet{newestqpo} that track QPO frequency.  We note that the time lags were measured at the QPO frequency.  Due to phase wrapping, the soft lag is expected to change signs and become a hard lag at $f=\frac{1}{2\tau_0}\sim2.5\times10^{-4}$ Hz \citep{pwrap,psd}.  We observe soft lags at the QPO frequency in four of the seven observations.  In each of these observations the best fitting QPO frequency was $<2.6\times10^{-4}$ Hz.  The remaining three observations with hard lags at the QPO frequency each had best fitting QPO frequencies $>2.6\times10^{-4}$ Hz.  This is consistent with phase wrapping occurring at $\sim2.6\times10^{-4}$ Hz, corresponding to a true soft lag of $\sim 1920$ s.  This soft lag corresponds to the light travel time across $\sim100$ $r_\mathrm{g}$.  We note that this radius matches the launch radius of the UFO discovered by \citet{rejUFO}.  Table~\ref{tab:info} lists whether a hard or soft lag was measured at the QPO frequency in each observation and the corresponding measured lag in seconds.

The intrinsic soft lag implies that the hard emission from the hot corona is reprocessed by the disk at all frequencies.  The increasing flux from the corona is expected to heat the disk and the Comptonized plasma spanning the disk surface.  As the disk heats, the seed photon temperature of the Comptonized plasma will increase.  The increased seed photon temperature and plasma temperature will increase the flux of the Comptonized disk emission, increasing the 0.3-0.5 keV flux.  We note that this overall effect is mild, given that the relative amplitude of the disk variability is quite a lot smaller than that of the coronal variability.

We can also measure the time lags at a particular frequency as a function of energy.  This lag energy spectrum is calculated by estimating the cross spectrum between a given energy band of interest and a broad reference band.  We use the full 0.3-10 keV band as the reference band and subtract the energy band of interest to avoid correlated errors.  We can compute the lag energy spectrum at the QPO frequency for each observation as well as the stacked lag energy spectrum at the QPO frequency.  The stacked lag energy spectrum is computed in the $2.6-2.9\times10^{-4}$ Hz frequency range, excluding the measured QPO frequencies of Obs 1-3 and 7.  This is done to avoid erasure of the shape of the stacked lag energy spectrum due to phase wrapping at $\sim2.6\times10^{-4}$ Hz.

\begin{figure}[htbp]
\centering
\includegraphics[width=\columnwidth]{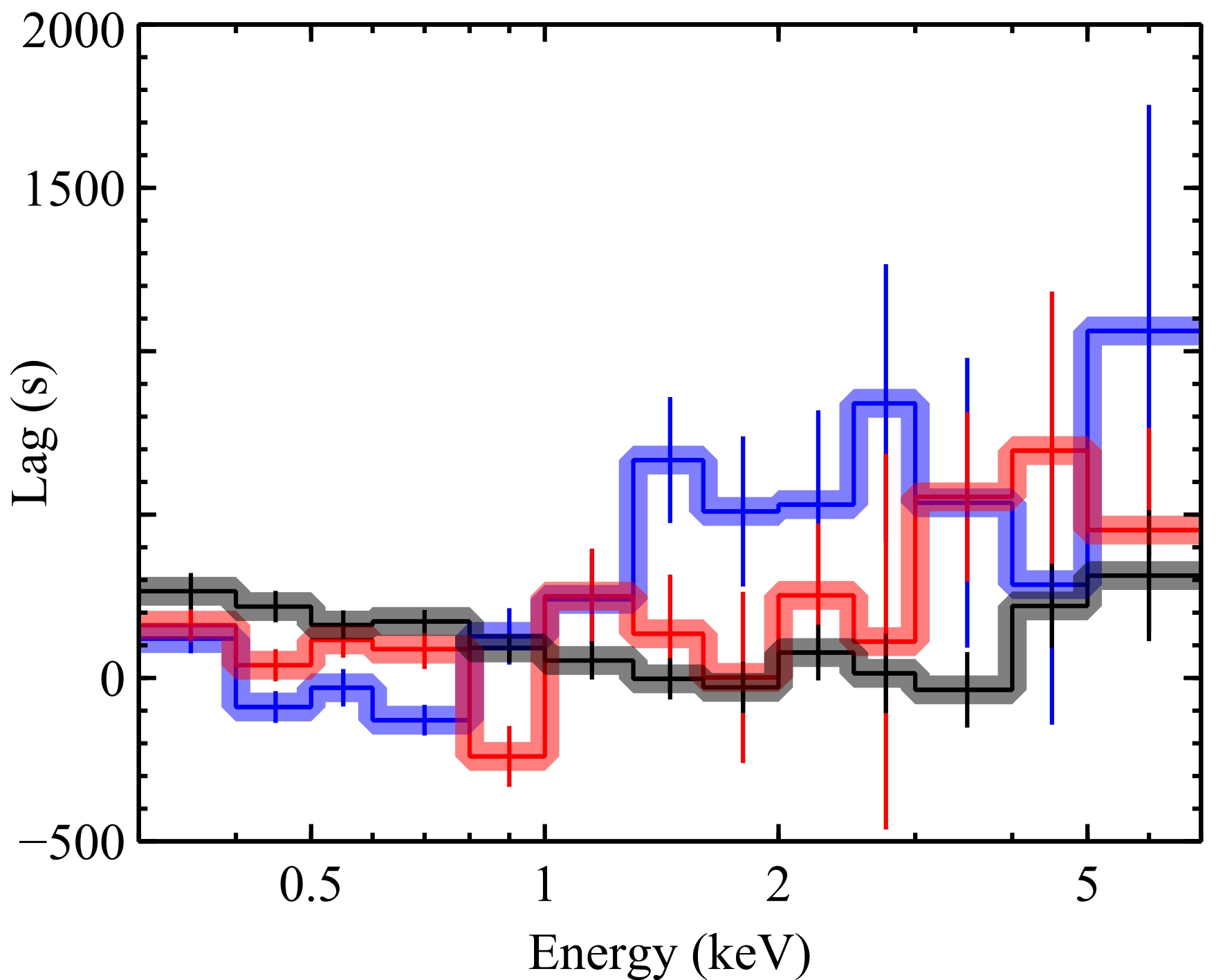}
\caption{The stacked lag energy spectrum of all seven observations in three frequency bins below the QPO frequency.  The frequency bins are $2-2.6\times10^{-4}$ (black), $1-2\times10^{-4}$ (red), and $0.1-1\times10^{-4}$ (blue). The soft Comptonized disk emission lag is superimposed on an overall hard lag that increases with decreasing frequency.}
\label{fig:lfLOW}
\end{figure}

The stacked lag energy spectra are consistent with models in which the soft Comptonized disk emission lags the continuum coronal emission (Figs.~\ref{fig:lf} and \ref{fig:lfLOW}).  Assuming that the soft Comptonized disk emission is responding to the coronal emission, we can predict the stacked lag energy spectrum based on the best-fitting model to the time-averaged X-ray spectra from the spectral analysis in \citet{rejUFO}.  The expected time lag is calculated by computing the average arrival time (accounting for relativistic time delays and all possible light paths) through the impulse response function weighted by the count rate in each time bin \citep{pwrap}.  We assume disk parameters from the spectral analysis in \citet{rejUFO}, impose a height to simulate the soft disk emission responding at disk radii of $\sim100r_\mathrm{g}$, and a reflection fraction to match the $R\sim1$ dilution.  We note that future work by Yu et al. (in prep.) will include more details of this model.  The expected lag energy spectrum closely resembles the stacked lag energy at the QPO frequency (see Fig.~\ref{fig:lf}b).  At the lowest frequencies, we observe the typical hard lag resulting from accretion rate fluctuations in the stacked lag energy spectra.  As frequency increases towards the QPO frequency, we see the soft Comptonized disk emission lag superimposed on the overall hard lag (Fig.~\ref{fig:lfLOW}).  Finally, at the QPO frequency we observe only the soft lag, now phase wrapped into a soft lead, and in remarkable agreement with the predicted lag energy spectrum at this frequency (Fig.~\ref{fig:lf}).  In some of the individual observations, it is possible to see phase wrapping in the lag energy spectra.  In these observations, the lag energy spectrum at frequencies just below $2.6\times10^{-4}$ exhibits the expected soft lag, while the lag energy spectrum at frequencies just above $2.6\times10^{-4}$ exhibits the likely phase wrapped hard lag.

\begin{figure}[htbp]
\centering
\includegraphics[width=\columnwidth]{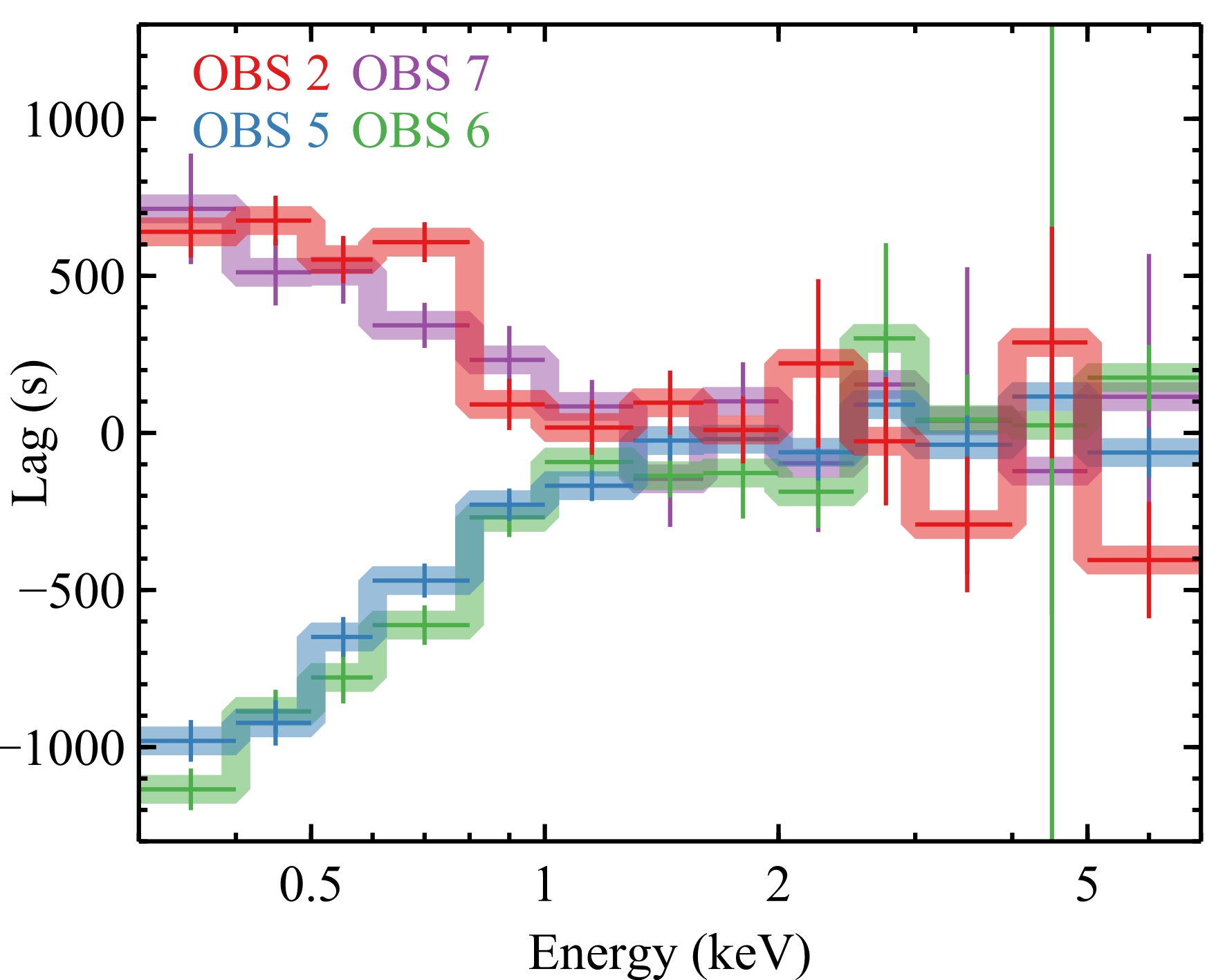}
\caption{The individual lag energy spectra of four of the observations.  Observations 2 and 7 show the soft Comptonized disk emission lag while observations 5 and 6 show the phase wrapped soft lag.  The slight asymmetry at low energies ($\sim700 s$ soft lag vs $\sim1000s$ hard lag) is due to the superposition of the low frequency hard lag shown in Fig.~\ref{fig:lfLOW} and commonly observed in AGN \citep{lags}.}
\label{fig:le_ind}
\end{figure}

\subsection{Covariance}
We use the covariance spectrum to investigate the energy dependence of the variability at different time scales.  The covariance spectrum is the cross spectral equivalent of the rms spectrum that measures the rms amplitude of variability as a function of energy \citep{cov}.  The signal-to-noise of the covariance spectrum is high because it uses a reference band light curve as a `matched filter' to isolate the correlated variations in each energy range, thereby rejecting any noise that is not correlated between the energy bands.  Calculating the covariance using \textsc{pylag }following \citet{psd} Equation (13) gives the spectrum in absolute flux units so that it can be interpreted like the time averaged X-ray spectrum.  Any spectral component that is just varying in normalization at the frequency of interest should appear the same in the covariance spectrum and time-averaged flux spectrum.

\begin{figure}[htbp]
\centering
\includegraphics[width=\columnwidth]{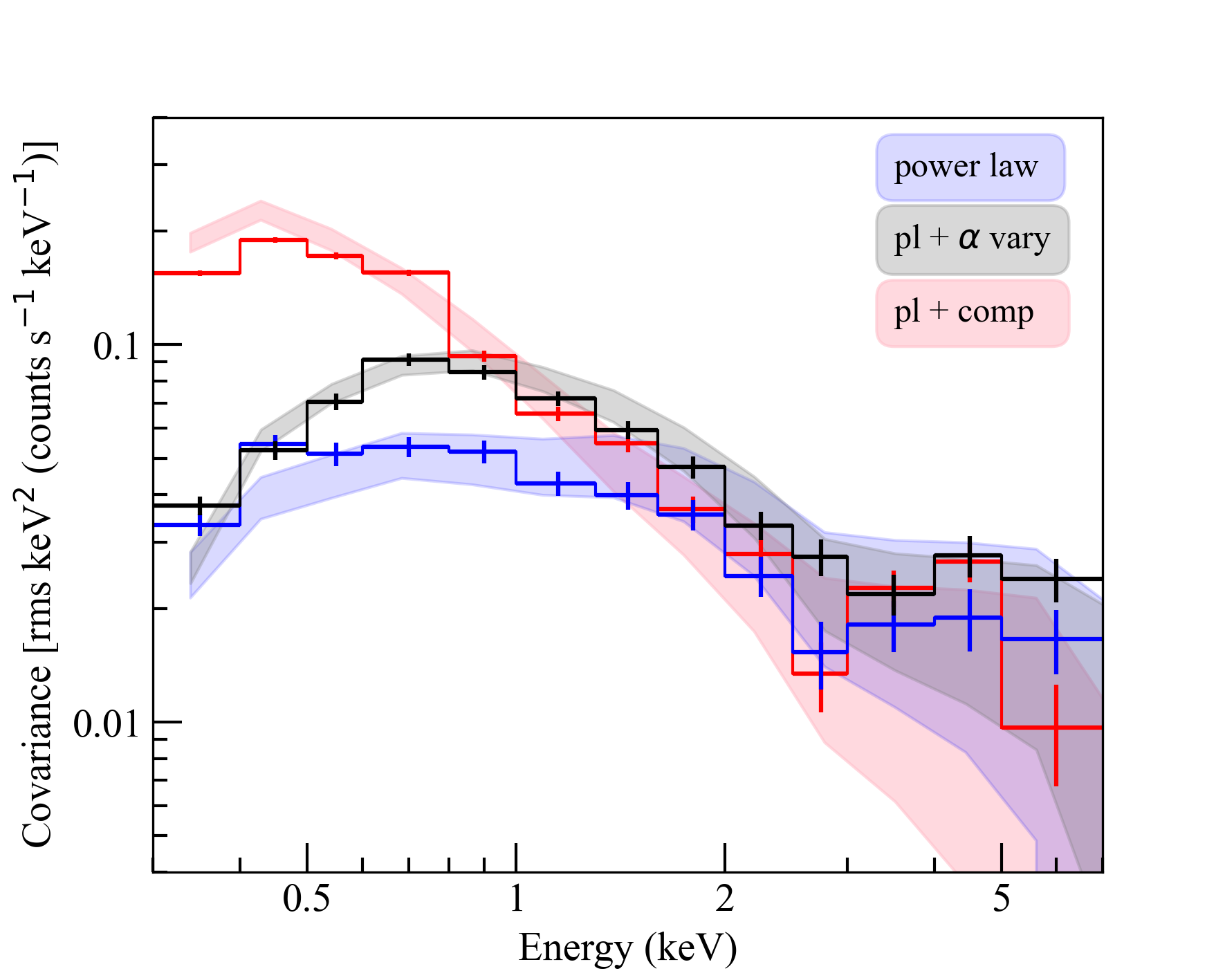}
\caption{The stacked covariance spectrum of all seven observations at low (red), QPO (black), and high (blue) frequencies.  Models for varying the normalization of the hot coronal power law like emission (light blue), photon index plus normalization of the hot coronal power law like emission (grey), and normalization of the hot coronal power law like emission plus Comptonized disk emission (pink) are shown.  The shading indicates the $1\sigma$ error on the simulated covariance spectra.}
\label{fig:cov}
\end{figure}

We investigate the covariance spectrum at the QPO frequency as well as at low and high frequencies either side of the QPO frequency.  These frequency ranges are $(0.5-1)\times10^{-4}$ (low), $(2.5-2.8)\times10^{-4}$ (QPO), and $(0.5-1)\times10^{-3}$ (high).  Fig.~\ref{fig:cov} shows the stacked covariance in each frequency range.  We again note that in each frequency range the covariance spectra are consistent across all observations, justifying the use of the stacked spectra.  At low frequencies, the covariance spectrum resembles the time averaged X-ray spectrum of RE\,J1034+396.  This indicates that there is variability in both the disk and the corona at low frequencies and that the dominant mode of variability is in the normalization.  At high frequencies, the covariance is lacking a soft excess.  This suggests that the soft disk emission is not varying significantly at high frequencies.

The covariance spectrum at the QPO frequency has a unique shape, peaking at $\sim0.7$ keV.  We simulate light curves accounting for galactic absorption and assuming that the coronal power law like emission has a time varying photon index and calculate the expected covariance spectra.  We find that the covariance expected due to power law like emission with photon index varying between 2.05 and 2.3 and its normalization set such that the flux at 0.175 keV is constant plus a component with power law like emission varying its normalization in phase such that the overall spectrum is softer when brighter resembles the QPO frequency coherence.  We note that the best fitting photon index to the time averaged spectrum of the seven observations is $2.15\pm0.02$.  The best fitting photon indices to the individual spectra span the 2.1 to 2.2 range \citep{rejUFO}, although we expect greater variability within each observation.  These results are again consistent with the QPO originating entirely within the hot corona.

We note that \citet{rejcov} infer that the covariance spectrum at the QPO frequency shows a soft excess with electron and seed photon temperatures $kT_e = 0.31\pm0.10$ keV and $kT_0 = 0.12\pm0.05$ keV, respectively.  These temperatures are much higher than the temperatures associated with the soft excess in their fit to the time-averaged X-ray spectra.  The power law with varying photon index model is consistent with the observed photon index variations from the time averaged X-ray spectra and does not invoke an additional spectral component.  The softer when brighter behavior is well motivated, as it has been observed in several other Seyfert galaxies \citep{SB}.  The varying photon index model is therefore the preferred model for the covariance spectrum at the QPO frequency.

\section{Discussion}

\begin{figure}[htbp]
\centering
\includegraphics[width=\columnwidth]{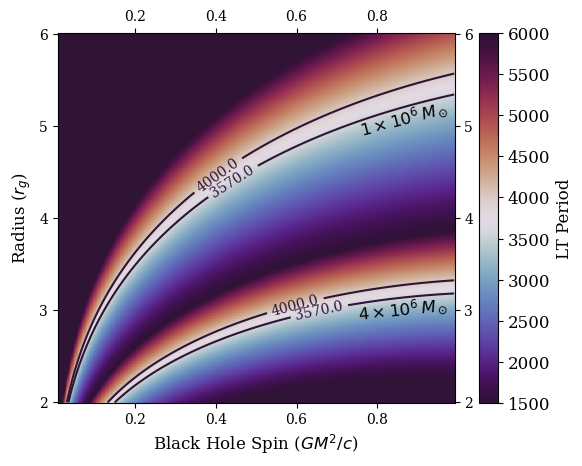}
\caption{Lense-Thirring precession period as a function of black hole spin and the radius of the hot inner flow for the precessing hot inner flow QPO model.  Contours indicating the upper (4000 s) and lower (3570 s) limits of the observed QPO period are indicated.  The upper band assumes the minimum mass, $M=1\times10^{6}M_\odot$, while the lower band assumes the maximum mass, $M=4\times10^{6}M_\odot$.}
\label{fig:LT_HIF}
\end{figure}

\noindent Given the evidence that the QPO in RE\,J1034+396 originates in the hot corona, we explore how this fits with several relevant models for the origin of QPOs.  One commonly invoked model for low frequency QPOs in BHBs is a precessing inner accretion flow.  This model assumes a truncated accretion disk with a hot inner flow misaligned with the accretion disk and black hole spin.  In the case of RE\,J1034+396, we consider the hot corona to be the hot inner flow.  In the simplest case, the QPO would arise due to the relativistic orbital frequency described in \citet{xrbrev}, given by

\begin{equation}
    \Omega_\phi = \frac{c}{2\pi r_\mathrm{g}}\frac{1}{r^{3/2}+a},
\end{equation}

\noindent where $r_\mathrm{g}$ is the gravitational radius, $r$ is measured in units of gravitational radii, and $a=J_\mathrm{BH}/Mcr_\mathrm{g}$ is the dimensionless spin parameter.  Given the possible mass ($M=(1-4)\times10^{6}M_\odot$) and black hole spin ($a=0-0.998$) values for RE\,J1034+396, the observed QPO frequency would constrain the hot corona radius, or equivalently the truncated accretion disk radius, to $r = 10-25$.  If the hot corona has a slightly elliptical orbit, modestly perturbed from the equatorial plane, we would expect periastron and Lense-Thirring precession.  Periastron precession is the rotation of the semi major axis of the orbit due to the inequality between orbital and radial epicyclic frequencies.  The Periastron precession frequency is given by

\begin{equation}
    \Omega_\mathrm{per} = \Omega_\phi - \Omega_r = \Omega_\phi(1-\sqrt{1-\frac{6}{r}+\frac{8a}{r^{3/2}}-\frac{3a^2}{r^2}}),
\end{equation}

\noindent where $\Omega_r$ is the radial epicyclic frequency.  If the QPO was due to periastron precession of the hot corona, the outer radius of the hot inner flow would be constrained to $r = 5-12$.  In such a system, Lense-Thirring precession would also occur due to the inequality between the orbital and vertical epicyclic frequencies.  The frequency of this vertical wobble is given by

\begin{equation}
    \Omega_\mathrm{LT} = \Omega_\phi - \Omega_z = \Omega_\phi(1-\sqrt{1-\frac{4a}{r^{3/2}}-\frac{3a^2}{r^2}}),
\end{equation}

\noindent where $\Omega_z$ is the vertical epicyclic frequency.  If the QPO was due to Lense-Thirring precession of the hot corona, the outer radius of the inner flow would lie within 6 $r_\mathrm{g}$, i.e. approaching the innermost stable circular orbit for a moderately spinning black hole (Fig.~\ref{fig:LT_HIF}).

\begin{figure}[htbp]
\centering
\includegraphics[width=\columnwidth]{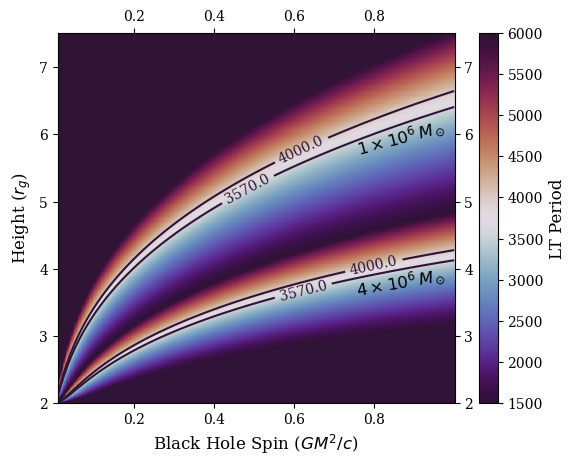}
\caption{Lense-Thirring precession period as a function of black hole spin and the height of the corona for the failed jet QPO model.  Contours indicating the upper (4000 s) and lower (3570 s) limits of the observed QPO period are indicated.  The upper band assumes the minimum mass, $M=1\times10^{6}M_\odot$, while the lower band assumes the maximum mass, $M=4\times10^{6}M_\odot$.}
\label{fig:LT}
\end{figure}

We can also consider the case where the hot corona exists as a failed jet.  In this scenario, the corona would sit above the accretion disk and have a spin axis that could be slightly misaligned with the black hole spin.  The misaligned spin would be expected to precess due to relativistic effects, another case of Lense-Thirring precession.  We consider a corona centered along the spin axis of the black hole at some height, $h$.  We make no assumptions about the spatial extent of the corona, but assume its orbit is misaligned with the black hole spin.  Under these conditions, the Lense-Thirring precession frequency is 

\begin{equation}
    \Omega_\mathrm{LT} = \frac{c}{r_\mathrm{g}}\frac{2ar\sqrt{\Delta}}{\rho^3(\rho^2-2r)},
\end{equation}

\noindent where we use the same conventions as before and $\rho^2=r^2+a^2$ and $\Delta=\sqrt{r^2+a^2-2r}$ \citep{LTcalc}.  Given the possible mass ($M=(1-4)\times10^{6}M_\odot$) and black hole spin ($a=0-0.998$) values, we use the observed QPO period to constrain the corona height to within 7 $r_\mathrm{g}$ (see Fig.~\ref{fig:LT}), consistent with expectations for typical corona heights \citep{corH,lags}.

\begin{figure}[htbp]
\centering
\includegraphics[width=\columnwidth]{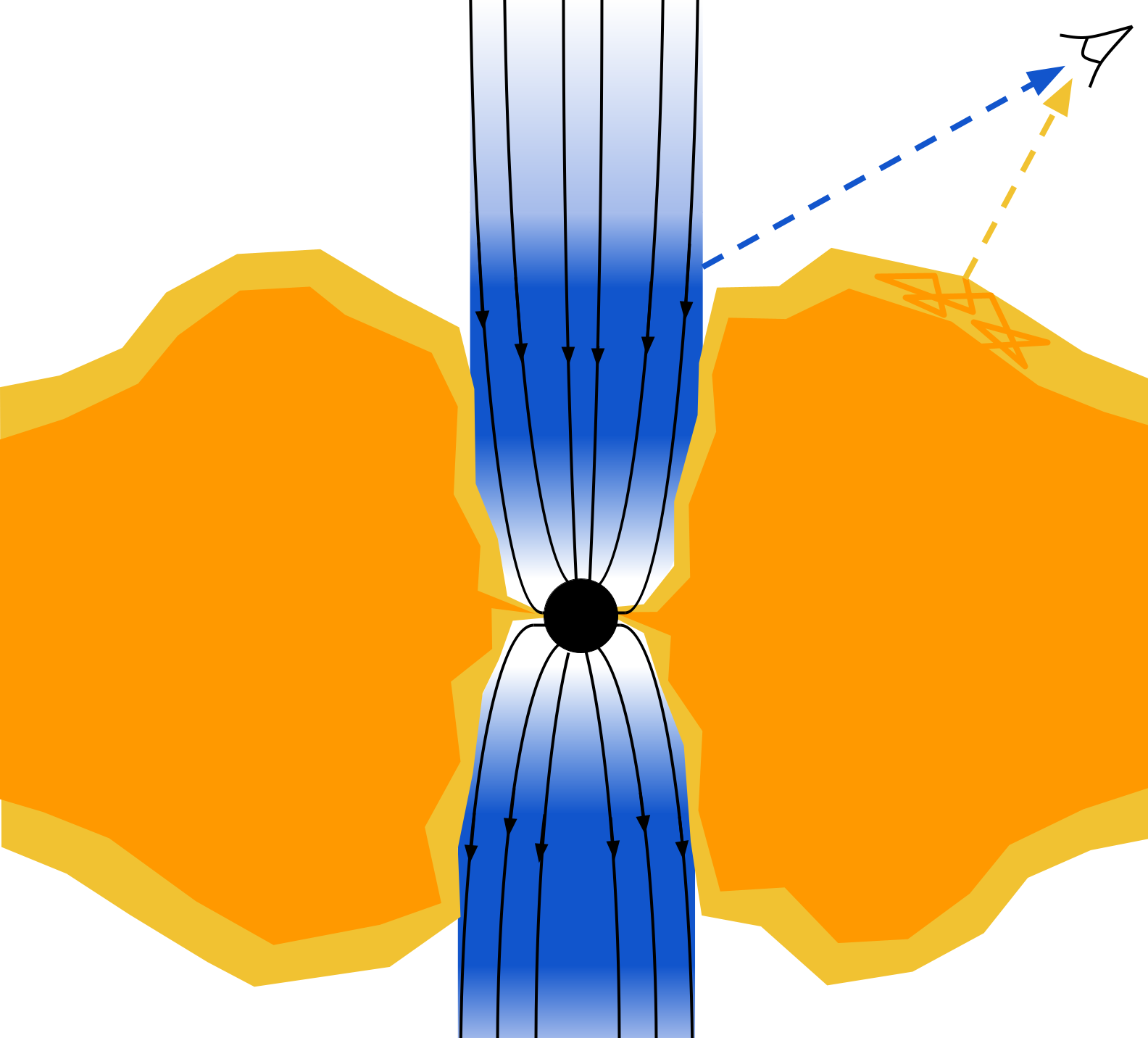}
\caption{Schematic of the inner regions of RE\,J1034+396 (in the MCAF scenario).  The turbulent accretion flow is illustrated in orange with the warm ($\sim0.2$ keV) Comptonizing plasma spanning the surface shown in gold.  Disk photons are Compton scattered to soft X-ray energies as they pass through the optically thick ($\tau \sim 13$) Comptonized plasma.  Accumulation of magnetic field (field lines shown in black) chokes the flow, separating the hot corona (blue) from the cooler, outer flow.  The hot coronal emission is responsible for the observed hard X-rays.  Magnetic interchange modes drive oscillations resulting in the observed QPO.}
\label{fig:dia}
\end{figure}

Finally, it is possible that RE\,J1034+396 hosts a magnetically choked accretion flow (MCAF) that drives the QPO, via the mechanism discussed by \cite{MCAF}.  In the MCAF scenario, the accumulation of poloidal flux creates a magnetospheric barrier in the inner regions of the disk, which separates the hot, inner corona from the cooler, outer flow (see Fig.~\ref{fig:dia}). Accretion through this barrier can still proceed, however, via magnetic interchange type modes, which drive oscillations at a frequency that depends on the angular rotation frequency of the black hole’s field lines at the disk-corona interface ($\Omega_\mathrm{F}$; the strong magnetic field at the barrier forces the plasma orbital frequency, $\Omega_\phi$, to be similar to $\Omega_\mathrm{F}$).  The field line rotational frequency depends on the black hole spin as 

\begin{equation}
    \Omega_\mathrm{F} \sim \frac{c}{2 \pi r_\mathrm{g}}\frac{1}{4}\frac{a}{2r_\mathrm{H}},
\end{equation}

\noindent where $r_\mathrm{H}$ is the radius of the black hole's event horizon in units of $r_\mathrm{g}$. Here, a factor of $\sim \frac{1}{4}$ comes from simulations where the disk-corona interface is observed at $\sim 4 r_\mathrm{g}$ \citep{MCAF}.  Based on the observed frequency of the QPO in RE\,J1034+396, and assuming the maximal mass of $4 \times 10^{6}$ \(M_\odot\), the MCAF model implies a spin for the central black hole of $a\sim0.5$.  For this spin, the disk-corona boundary corresponds approximately to the innermost stable circular orbit.  In this case, one might also anticipate oscillations due to periastron or Lense-Thirring precession.  In this scenario, the periastron frequency happens to be very similar to $\Omega_\mathrm{F}$, while the Lense-Thirring precession frequency will be lower ($<1\times10^{-4}$).  Unfortunately, the existing X-ray observations of RE\,J1034+396 are too short to robustly detect a QPO signal at the Lense-Thirring precession frequency.  Within the MCAF picture, the enhanced magnetic flux of the inner accretion flow may also help to energetically stabilize the warm ($kT\sim 0.2$ keV) optically thick ($\tau \sim 13$) Comptonized plasma that spans the disk surface \citep{warmC}.

\section{Conclusions}
\noindent We have re-examined the origins of the QPO in $\sim1$ Ms of publically available \textit{XMM-Newton} observations of the nearby narrow-line AGN, RE\,J1034+396.  Given the refined spectral modeling presented by \citet{rejUFO}, we are able to (relatively) cleanly separate the disk (0.3-0.5 keV) and hot corona (2-7 keV) emission components, with the soft band having only $\sim10\%$ contamination from the hot corona.  
Our analysis of the periodograms from these two energy bands indicate that the QPO signal in the soft band can be entirely attributed to contamination from the hot corona, providing strong eveidence that the QPO arises from processes only involving the hot corona.  The peak in coherence at QPO frequency aligns nicely with this interpretation.  The covariance spectrum supports the interpretation that the QPO is isolated to the hot corona, being well described by fluctuations in the coronal power law photon index.  We note that the phase wrapping interpretation implies that the soft disk emission lags the hot coronal emission, indicating that some variability in the hot corona is reprocessed by the disk at all frequencies.  Given the evidence that the QPO arises solely in the hot corona, models that assume origins in the accretion disk are disfavored.  We consider several models that could generate the QPO in RE\,J1034+396 and consider their implications.  Depending on the model, we are able to constrain the radial extent of the corona, the height of the corona, or the black hole spin.  RE\,J1034+396 is one of only two AGN with compelling evidence of a QPO.  This is the first time such a QPO has been convincingly isolated to a single spectral component.

%If we adopt the precessing inner accretion flow model, where the corona is the hot inner flow, and assume Lense-Thirring precession of the corona produces the QPO, we can constrain the radius of the inner flow to within 6 $r_\mathrm{g}$.  If instead we model the corona as a failed jet with misaligned spin some height above the accretion disk, assuming Lense-Thirring precession of the corona is responsible for the QPO, we can constrain its height to within 7 $r_\mathrm{g}$.  

\begin{acknowledgments}
We thank Zhefu Yu, Robert Wagoner, and Roger Blandford for helpful discussions. This work was supported by the NASA Astrophysics Data Analysis Program under grant number 80NSSC22K0406. CST is supported by the ABB Stanford Graduate Fellowship.  SWA is supported in part by the U.S. Department of Energy under contract number DE-AC02-76SF00515.
\end{acknowledgments}

\vspace{5mm}
\facility{XMM-Newton}

%% Similar to \facility{}, there is the optional \software command to allow 
%% authors a place to specify which programs were used during the creation of 
%% the manuscript. Authors should list each code and include either a
%% citation or url to the code inside ()s when available.

\software{XSPEC \citep{xspec}, \textsc{pylag} (\url{https://github.com/wilkinsdr/pylag})}

\bibliography{ref}{}
\bibliographystyle{aasjournal}

\end{document}